# Three species of neutrinos
# Six flavors quarks and three gluons.
# (Consequences or coincidences?)

Alexandru Mihul, Eleonora Mihul
Bucharest University
Mailing address Alexandru.Mihul@cern.ch

**Abstract.**
A linear real 4-dimensional space structured by Lorentz metric implies six 2-dimensional singular subspaces and three independent 1-dimensional subspaces called radicals. If we consider meaningful to use them as representing six quarks and three gluons, respectively, then antiquarcs and antigluons are implicitly defined. Since Lorentz metric is not singular, the dual metric space of X denoted in general by X* is well defined and follow the structure of X. The peculiar properties of singular subspaces guaranty what we call confinement. For precision, we define first what means free system using the structure of X and X*and thereby how X are lodging six quarks and three gluons.

**Introduction.**

Quantum chromodynamics is relied on the relation between length and momentum in quantum mechanics including so the hypothesized short distance interaction of the quarks inside a nucleon. Relativistic mean field description uses for instance the infinite momentum frame. The aim of this paper is to show how length and momentum are implied by a space structured by Lorentz metric X and how six quarks and 3 gluons can be represented in such a space.

. In the sec. 1, the relation between energy and length is obtained as consequence of Lorentz causality. It means that for any certain energy of a free system there is a length so that their product is equal to the causal constant L For rest (massive) system and for the photon - restless (mass less) system the product is L.

In sec. 2 we show how number three of neutrinos species is also the consequence of Lorentz causality and for neutrinos species energy multiplied by length is L/2 instead of L.

In sec. 3 we use the six 2-dimensional singular subspace of X and three 1-dimensional subspaces of X in which six quarks and three gluons can be represented.

Finally we sketch in sec. 4, how electric charge and so electric (and gravitational) interaction is implied by the Lorentz causality for understanding the interaction between quark electric charges.

## 1. Causal constant and "Asymptotic freedom"

Lorentz interval divides the space X in causally related points, - i.e. ordered - on the two light cones and on the boundary of the cones [1], and not causally related points, it means space-like related points.

It is important for what follows that causal relation [2] is an ordering relation of X with regard to which X is directed [3]. It means that for any x' and x'' in X, there is an x''' in space X so that x''' is causally related to x' and to x''. If X is an ordered linear space, then X is directed if and only if the cone generates X, the Clifford theorem [4].

Certainly, the structure of X by itself cannot provide properties of a physical system.

However, significant clue to the problem is that the Lorentz metric is not singular, (unlike of the Newtonian affine space). So its dual metric space is precisely defined. The components of a four-dimensional vector p of the dual space X*, are the coefficients of the real-valued linear functional

$$p(x) = p_0 x^0 - \mathbf{p}.\mathbf{x} \quad \text{(bilinear form)}. \qquad 1)$$

To make short the writing we denoted by x, a "vector", x'-x''. Hence, we are factually dealing with two dually "related" spaces X and its dual, X*.

The invariance of p(x) under the translation group, subgroup of the Causal Group [5], provides a structure defined by the following constraints interpreted as describing a free physical systems defined by 1):

I.  a certain 4-dimensional vector p,
II. x belonging to a non extendible causal curve in X (ray is a particular case);
III $p^2 = Y$, $p^2 = p_0^2 - |\mathbf{p}|^2$, the invariant Y to the dual causal group is positive or zero;
IV. the invariant $Y = 0$ i.e. $p_0 = |\mathbf{p}|$ if and only if x belongs to the ray i.e. $x^0 = \pm |\mathbf{x}|$

**Definition.** That system (particle) representable by such a structure, we call Free Physical System (FPS).
If IV is also satisfied, we call rest-less FPS.

**Comments**. Empty space does not exist, neither the ideal stable vacuum. They are, like free particles, an idealization. Thus, the isolated system is an intellectual challenge and so its properties can be provided only as consequences of the "principle" we start with. But let to cite the beautiful suggestion of Newton: "Haven't the small Particles of Bodies certain Powers, Virtues, or Forces, by which they act upon one another for producing a great part of the Phenomena of Nature?"

**Ordered (Causal) Limit.**

Mathematically, the constraints I - IV mean that a certain p maps, an ordered sequence p(x) of a non extendible causal curve, in real numbers R.

According to E. H. Moore and H.L Smith [6], each of such an ordered set of numbers p (x) for x defined on a ordered set in a directed space with respect to the ordering (causal) relation has a limit L.

Now, Lorentz causality excludes zero value for this limit, orthogonal condition of a Lorentz metric (it is zero for a space like vector-hence not causal), but it can be of any smallness. Therefore, one proves easily [7] that for any p, one can choose the same real limit for L. Then, we have

$$p_0 \chi^0 - p_1\chi^1 - p_2\chi^2 - p_3 \chi^3 = L \qquad 2)$$

**Definition**. This small value L, we call ordered Lorentz (Causal) Limit

It means that for any given p with either $p^2 > 0$ or with $p^2 = 0$, there is a certain vector $\chi$ so that 1) is satisfied.

Important thing is to attach a physical meaning to L. For that, we have to choose measurable entities to the components of x and to the components of p.

Surprise or it is trivial? Since the Lorentz interval is a sum of the components (squared) of x, it is obvious that all components must represent the same entity. We chose length for all four components of x and, for all components of p we choose energy. Time is an outside parameter that provides the constant c by using experimental data. It represents a limit that defines the boundary of the cone.

Then, the dimension of L is length multiplied by energy (just the dimension of the square of electric charge as we shall get in the last section).

In our mind's eye we find something that has meaning to us, but unless we have objective proof of checking it, we don't know if "the something" has any meaning. Therefore, the justification of Lorentz causality, even if it is motivated as representing inductive logic, rests in the proof of usefulness for physics [5].

Now, we have to find those properties of the FPS that are direct consequences of the detailed structure of X and X*, and so expressed through the Lorentz Limit. L

For $p^2 > 0$ rest ("massive") particle, **p=0** being compatible with the above constraints we find that $\chi^0$ represents a natural scale

$$\chi^0 = \pm L / p_0 \qquad 3)$$

attached to any such a particle with a certain energy $p_0$. Confronting with empirical data, we find that we got the scale attached to a rest ("massive") particle, just what we call Compton wavelength $\lambda = h/mc$ for $p_0 = mc^2$ and so:

L = hc, is a physically causal constant.

**Consequence**. Planck constant h is related to Lorentz causality as much as c is.

**Note**. As it is known, the wave description of matter defines a natural scale for a particle through its Compton wavelength. The relation 1) shows that for a space endowed with Lorentz metric, the constraints imposed by Lorentz causality imply such a scale without any other assumption.

We use rest and restless instead of massive and massless, because having not defined time we cannot define c as a speed and neither mass.

**Remark**. It is outstanding that the above mathematical statements, structures, are valid either for $x^0 > 0$, or $x^0 < 0$ (the two cones) referring to L positive or negative. This provides the concept of system and anti-system, in particular particle and antiparticle regarding to the relation to the space.

As we do not have any other assumption, the meaning or meaningless of the consequences can be seen as true or false of Lorentz causality, obviously within their sphere of applicability.

**2. Restless systems.**

Now let us analyze the "massless" or better call them restless systems for which $p_0 = |\mathbf{p}|$ and $x^0 = \pm |\mathbf{x}|$. In this case, clear that only the relative orientation between **p** and **x** is left to be chosen. The exquisite tandem between these two "vectors" is the clue of the "dynamical" structure of the restless systems.

There are two invariant possibilities:

One is transverse, i.e. 3-dimensional **p** and **x** are orthogonal. This is what we call rest-less transverse FPS. For this case, (the scalar product being zero in 1)) we have:

$$p_0 \chi^0 = \pm L \qquad \qquad 4)$$

which is just the blackbody Planck relation for energy of the photon. The handedness -L and +L represent two signs of "helicity" (spin). The norm of 3-dimensional **p** is just the energy of the system. However, as a vector, **p** has to be interpreted as angular momentum indicating the both possible rotations with respect to the ray. This provides both orientations of the closed paths and so the elements of the first homotopy group of U(1) space group. Using the fact that the finest topology of X, induces a discreet topology on the ray [8], the winding number of homotopy group "measures" the distance "covered" by the photon and so, infinitely connected topology allowed photon to be stable "forever" and topologically being represented by a cylindrical space. Is this the root of U(1) symmetry of the photon?

It is remarkable that the relation 3) and 4) are formally the same.

Another rest-less system is "longitudinal", i.e. **p** and **x**, for $x^0 > 0$ must be antiparallel, pointing in opposite direction, left handed, "helicity" equals $-1/2L$ according to 2). For $x^0 < 0$, they must be parallel with $+1/2L$, right-handed antineutrino and

$$p_0 \chi^0 = \pm L/2 \qquad \qquad 5)$$

Thus a system, as free one, with properties of which we call neutrino and antineutrino follow directly from 2). Obviously that the energy is positive for neutrino as it is for, so called, antineutrino.

Thus the change of the sign of energy for definition of particle and antiparticle is off..

**The energy-length rule.**

Hence the Lorentz causal relation defines three classes of free isolated systems:
a class of massive ones for which we have the relation:
$$P_0 x^0 = \pm L$$
and two classes of massless systems:
transversal for which we have the relation
$$p_0 x^0 = \pm L$$
and longitudinal systems for which
$$p_0 x^0 = \pm L/2$$
For massive system, $x_0$ provides what we call Compton wavelength. and for massless photon its wave length

Now let us discuss the subtle issue of dimensionality of the space X lodging simultaneously massive systems, one species of transverse mass less and. three [9] species of longitudinal systems, photons and neutrinos. For each system being defined its antisystem.

The transverse systems being defined by the outer product, the space X must be at least 4-dimensional. Instead the exquisite role in describing longitudinal system, beside mass less, is played by the inner product .**p.x**. This lowers the grade of a vector so can perfectly be defined in 2-dimensional space. So our space X were 3-dimensional, if longitudinal restless systems there would be only. Since we have to use the same space, as it is implicated in the way we introduced the concept of the space as representing Lorentz causality (Partially ordered structure), longitudinal system that requires only the inner product must be represented in a 3-dimensional subspace of X .

Then the quest is, how many linear independent 3-dimensional subspaces we have. The space X has seven 3- dimensional subspaces. One is 1-dimensional space like, three are time-like 3- dimensional subspaces and. three are 3-dimensional singular subspaces [10]

Except time like subspaces can lodge longitudinal systems, with all their properties being restless and longitudinal. Since there are three such subspaces there should be three diverse types of "neutrino" and obviously their antineutrino.

**Some details**:
X is generated by a basis of one time-like and three space–like vectors...
If we keep the time like basis vector (1,0,0,0) and choose one of the three possible combinations for instance (0,1,0.0) and (0,0,1,0), we get the basis that generates one of three, 3-dimensional time-like sub-spaces. In each of them, independently we can define the restless longitudinal system.

The 3-dimensional space-like subspace generated by three space-like basis vectors is excluded "by causality."

Peculiar are three singular spaces whose basis include one light-like vector, like (1,1 0,0), which together with (0,0,1,0) and (0,0,0,1) generate a singular 3-dimensional singular subspace. Rather akin to the longitudinal system, they could play a exquisite role in the neutrino mass problem.

Obviously there are three such singular 3-dimensional subspaces [10]

**Partner**.

If we are looking now for a partner of neutrino, we have to find a system whose some properties are common.. Symptomatic for neutrinos is a certain sharp handedness. Let say left for neutrino and right for antineutrino. and they are representable in a time-like 3-dimensional subspaces  Obviously that the partner, if exist must be a rest system otherwise would coincide with neutrino itself.

According to the above finding; using the same energy-length law, we shell show that the curl of the negative charge, electron, is strictly left and positive for positron. That is why neutrino "chooses" electron as partner and antineutrino chooses the positron. Thereby electron must be "particle" like neutrino and positron "antiparticle" like antineutrino.

Then the question is: does the above structure imply the existence of such rest-systems?

**Remark** about commutator and anticommutator. If we define now causality on the 2x2 matrix representation of the SL (2, C) group, we find that the matrices corresponding to **p** and **x** anticommute for longitudinal case and, commute for transverse case. The multiplying factor of unit matrix is just causal limit L according to 4)

**3. Quarks.**

There is a pertinent question. If so many properties of the free systems can be obtained from the structure of the spaces X and X* and their concatenation, can we find that the number of  so called "quarks" whose basic property is just that they do not exist in X as free particle, defined in 1), must be six,?

So we start with the above structures.

q1) there are six singular two dimensional subspaces of X, if they may represent six quarks. Then.

q2)  there are six antiquarcs (according to the above definition, using two cones),

q3) confinement. Precisely, since the magnitude of $p^2$ of a vector in any singular space is negative they cannot be free, since the condition III is violated.

q4) instead of a linear combination of such vectors provides vectors for which $p^2$ is positive, and so providing the real particles.

q5 ) Now if we accept to call fermionic, the property described by scalar product between 3-dimensional x and p ,which in matrix representation is expressed  by anticommutator then quark has fermionic property since the structure implies such a structure.

Now let us use the properties we enumerated for neutrino.

The analogy with the 3-dimensional singular subspaces of X, that can be used to get the mass for neutrino (for the longitudinal systems), suggest the following possibility to get the above properties of quarks

Precisely, each 3- dimensional singular space has two, 2- dimensional singular subspaces. Indeed for our example one is generated by (1,1,0.0), (0,0,1,0) and the other by(1,1,0.0), (0,0,0,1) ("up" and "down").

The classification of the six subspaces of X includes also six singular 2-dimensional subspaces in the dual space. As we already mentioned the singular subspaces have a peculiar properties. On one side they are related to our world through the cone by their radical subspace. The space X implies three radicals one dimensional subspaces (subspace whose all vectors are light-like) each being generated by (1,1,0,0,), (1,0,010,), and (1,0,0,1,) respectively. On the other side since the invariant Y (the mass) of the vectors in the dual space, $(p_0)^2 – (p_1)^2 – (p_2)^2$ being negative because, $(p_2)= (p_1)^2$, III is obvious not satisfied. Instead a linear combination of them, of different flavor, provides the masses of the real particles.

That is why quarks cannot appear like a free system. So, if the six singular 2-dimensional subspaces of X are used for describing six quarks then confinement is provided by very structure of the Lorentz space and thereby the mystery of confinement is off.

Interesting that in chromodynamics the asymptotic freedom is conditioned by the number of distinct quark flavors due to the fact that the coefficient of the lowest order term in the perturbation expansion of the beta function is proportional to $–(11/2 – n/2)$, where n should be the number of quarks.

If the space X is responsible for three neutrinos and only three, and also for six quarks and only six, let us try to find another property of quarks related to X. Is electric charge one of them?

**4. Two point system.**

Since electric charge plays an important role in the supposed interaction between quark using as usually Coulomb law we should like to scathe in this section how charge and so Coulomb law are consequences of Lorentz Causality.

Obviously, Lorentz duality cannot distinguish a massive free "charged" system, whatever we call charge, from a "neutral" one. This is the reason why we can consider them, charged or not, as free, isolated systems representable on the points on a causal curve.

Now we put the following problem. Is it possible, using the above way of reasoning, to find some properties of an entity which we call charge, carried by two massive system represented in two points x' and x'' with finite interval i.e. Int.(x'-x'') = finite.

**We call it two-point system.**

So let there be such a class of free systems whose representation of any of them, as isolated, requires two and only two points, x' and x'' in X,. At each point x' and x'', there exist an entity denoted by Q, whose properties related to Lorentz duality, we are looking for.

We assert that the entities Q are manifested through one kind of communication.

There must be one kind of communication between the two entities; otherwise they would be an isolated free system, according to the definition of isolated systems represented in points of a causal curve.

Now, we suppose that the effect of communication is manifested by energy that we denote by $A_0$. Hence, $A_0$ must be a function on the entity Q and on the interval Int. (x'-x'') and, obviously on p (or m).

That is to say:

$A_0 = F(p, Q, Int.(x'-x''))$.

We drop for the moment the dependence on p; it will appear naturally as it has to be expected.

We discuss now the simplest possible form of function F.

For that, we assume (rather a strong assumption) that F is factorized. It means

$$A_0 = f(Q) g(Int(.x'-x'')) \qquad 6)$$

To be relevant to physics, that's to say compatible with the consequences of Lorentz duality, the points x', x'' and the interval {Int.(x'- x'')} must be in the causal part of X. Then, we have to choose one of the two possibilities:
Either
$$Int.(x'-x'') > 0$$
7)
or
$$Int.(x'-x'') = 0 \qquad 8)$$

Let us consider that the interval is light like 8) and denote by x = x'-x''. Then g is a function only, say, of zero component of x. So, we have

$$A_0 = f(Q) g(x^0) \qquad 9).$$

We stress that the properties of entity Q, we are supposed to be looking for, are those that follow from embedding them in X. Therefore, the energy is to stand up the energy-length rule. In other words, there is $y^0$ for any value of $A_0$ such that $A_0 y^{0.} = L$ so that:

$$f(Q) g(x^0) y^0 = L \qquad 10)$$

Since L is constant the left side of relation 10) has not to depend on $x^0$. This can be so only if $g(x^0) = 1/ky^0$ where k is just a dimensionless constant of proportionality..

Then, for the light-like interval of x, $x^0 = 1/ky^0$ fit in, uniquely, with the energy-length rule.

Substituting $x^0$ by r (using 8)) we recovered a law of proportionality between energy and inverse of the Euclidean distance r between x' and x''.

$$A_0 = f(Q) / kr \qquad 11)$$

**Remark. 1**. As we relied our result on causality, that implies that zero value for $x^0$ is physically meaningless. The same is true for the distance r. And this, in its turn, defines two spheres in X localized at each point x' and x''. Thus, the communication depending on r expresses a locality (continuum) property provided by the superposition of the two spheres with the center in x' and x'', with radii r being Euclidean distance between two points. Hence, the entities we are looking for, implement a principle of locality which Faraday called field, and it is clear now why Coulomb and Newton laws can be treated as classical ones. But, it cannot find why F (Q) up to a dimensionless constant k is just L = hc.

**Remark. 2**. The constant $k = x^0/y^0$ is actually the proportional constant between the scales defined by free systems, and that, which fits for Q.

We note that the fine structure constant alpha is the quotient between electrostatic energy of two systems of the same mass, each carrying a unit of electric charge at the distance of their Compton wave length, and the proper energy of p or m. It is obvious a dimensionless constant. That is what we have got.

Let us go further. We actually got a dual 4-dimensional vector $A = (A^0, \mathbf{A})$. With this, one defines a bi-dimensional tensor. **This,** in its turn, is decomposable in two invariant anti-symmetric which we denote by $F^{lm}$ and, $T^{lm}$ the symmetric tensors.

For anti-symmetric one, two signs of the charge Q must exist to realize the positive and negative sign of curl. **A**.

Now, the f (Q) cannot be the sum, i.e. f (Q) = 2Q, because for different signs of charges it would be zero. So, the only choice is $f(Q) = Q^2$ and so, $Q^2 = \alpha$ hc.

**Remark. 3**. The vector potential **A** is uniquely defined, as it is actually proved by Aharonov - Bohm experiment. It is the operator curl, or co-boundary operator, which by its property is defined up to an arbitrary co-cycle.

**Remark. 4.** Now, the meaning to introduce canonical momentum p + A for a charged particle in the Dirac equation is evident. Both p and A are two vectors in the dual space.

**Remark. 5**.The Maxwell equation is written as:
D F = 0
which means that the 2-form F satisfies the above equation, where D is the co-boundary operator or F = D A. The second Maxwell equation is obtained by introducing the Hodge dual what is well known.

**Remark. 6**. The conservation law of electric charge and the quantification of electric charge are embedded in the above properties. The dual role of electric charge as a conserved quantity, charge and anti-charge, and the fact that it also measures the strength of the communication is obvious.

Now the question is, which is the space where takes plays the interaction between quarks?

**Newton law, f(p).**

In this case, we cannot have a constant of proportionality, because $y^0$ was introduced for the particles of any charge. It represents a scale, which depends on the mass and only on the mass. Thus, a constant of proportionality equivalent to alpha is meaningless.

In this case, important is that the symmetric tensor $T_{lm}$ in the dual space appears naturally. Then it has a correspondent in the space X, $R^{lm}$ that implies a curved space. I think it is extremely important that even for the simplest kind of interaction, two-point isolated system, and the space X is no more a flat space.

**The interpretation of Newton law of gravity** defined for the mass m located in each point x' and x'' reveals straightly what is called, Planck mass. The concept of Planck mass was introduced as the mass for which the Compton wavelength is equal with half of the Schwarzschild radius. Thus, $L = hc$ appears naturally related to the gravitational field.

*Conclusions*

There is no one event, which can be grasped in the frame of causality without implying L. The approximation implied by causality can also be related to how many causal mappings are taking and this also measures the departure "from flatness" [5].

As we have seen, Lorentz causality constraint implies a scale for free systems given by zero (length) component of x. For rest system (massive) and restless (massless) transversal system (photon), the scale is the same, i.e. for the same energy of two different systems; the ratio of the scales is 1. Instead, the ratio of the scale of a restless transversal (photon) and of a restless longitudinal (neutrino) at the same energy is 1/2. And the electrostatic scale is given by the alpha constant.

When we consider instead of linear $R^{1+3}$ a manifold, we assume that the metric tensor field g(x) with its associated metric connection (expresses the definition of parallelism-parallel transport-implied by the metric) is the only geometrical structure possessed by the manifold. The square matrix g(x) is symmetric. Hence, in any given point, in a coordinate system, we can adjust the scales of the coordinates so that each element of g is either +1 or -1. If the equivalence principle is to hold in the neighborhood of the point, then one value must be +1 (or -1) and the three others -1 (respectively +1). But it seems that only in a very small region of a point of a manifold it is meaningful to choose the flat space with Lorentz metric. However, regardless of the richness of structure, it can only provide a "basic" geometry. The straight ray becomes a respective geodesic.

Then what is structure of the space in which the interaction of quarks must be imbedded?.